\documentclass[runningheads]{llncs}
\usepackage{graphicx}
\begin{document}
\title{Exploiting bilateral symmetry in brain lesion segmentation}
\titlerunning{Exploiting bilateral symmetry for brain lesion segmentation}
%

\author{Kevin Raina \inst{1} \and
Uladzimir Yahorau
\inst{1} \and
Tanya Schmah
\inst{1}}

\authorrunning{Raina et al.}
%
\institute{University of Ottawa, 585 King Edward Ave, Ottawa, ON K1N 6N5\\
\email{\{krain033, uyahorau, tschmah\}@uottawa.ca}}

\maketitle              
\begin{abstract}
Brain lesions, including stroke and tumours, have a high degree of variability in terms of location, size, intensity and form, making automatic segmentation difficult. We propose an improvement to existing segmentation methods by exploiting the bilateral quasi-symmetry of healthy brains, which breaks down when lesions are present. Specifically, we use nonlinear registration of a neuroimage to a reflected version of itself (``reflective registration'') to determine for each voxel its homologous (corresponding) voxel in the other hemisphere. A patch around the homologous voxel is added as a set of new features to the segmentation algorithm. To evaluate this method, we implemented two different CNN-based multimodal MRI stroke lesion segmentation algorithms, and then augmented them by adding extra symmetry features using the reflective registration method described above. For each architecture, we compared the performance with and without symmetry augmentation, on the SISS Training dataset of the Ischemic Stroke Lesion Segmentation Challenge (ISLES) 2015 challenge. Using affine reflective registration improves performance over baseline, but nonlinear reflective registration gives significantly better results: 
an improvement in Dice coefficient of
13 percentage points over baseline for one architecture and 9 points for the other.
We argue for the broad applicability of adding symmetric features to existing segmentation algorithms, specifically using nonlinear, template-free methods.


\end{abstract}
\section{Introduction}
Segmentation of stroke lesions and tumours in neuroimages, also called lesion mapping, can give valuable information for prognosis, treatment planning and monitoring of disease progression. 
The ``gold standard'' for lesion segmentation is still manual delineation by a human expert, going through each of the horizontal slices of the three-dimensional image and labeling each separate voxel as either healthy or belonging to a lesion. This is tedious, time-consuming, and often impractical, and therefore in practice, a human expert usually gives only a qualitative assessment of lesions. Further, there is inter-observer variability; the size of this variability varies significantly by task, but we note that an average Dice score of 0.58 for overlap of manually-outlined lesions by two raters was reported for the ISLES2016 challenge \cite{winzeck2018isles}.  These observations indicate a need for automatic brain lesion segmentation algorithms. However, accurate lesion segmentation is a challenging task for many reasons, including large variability in location, size, shape and frequency of lesions across patients.

While a plethora of automatic lesion segmentation methods has been proposed, most of the currently leading methods are based
on convolutional neural networks (CNN)\cite{winzeck2018isles}.
Many of these use 2D CNNs, where the 3D neuroimage is processed as a sequence of independent 2D slices. It is worth noting that these approaches are arguably suboptimal, since they do not take into account the 3D spatial structure of the data. 
Nonetheless, many 2D methods have shown promising results, including the methods of Havaei et al.  \cite{Havaei2017brain} and Kamnitsas et al. \cite{kamnitsas2017efficient},
which we use as baseline architectures in the present paper due to their limited memory requirements.
Some other works used CNNs with an input of three orthogonal patches around each voxel being classified, thus incorporating some 3D information, however this significantly increased memory requirements and computational complexity. The technique of dense inference greatly sped up inference time, and led to several successful 3D segmentation methods, see discussion and
references in Kamnitsas et al. \cite{kamnitsas2017efficient}. 

\begin{figure}
	\centerline{\includegraphics[width=\textwidth]{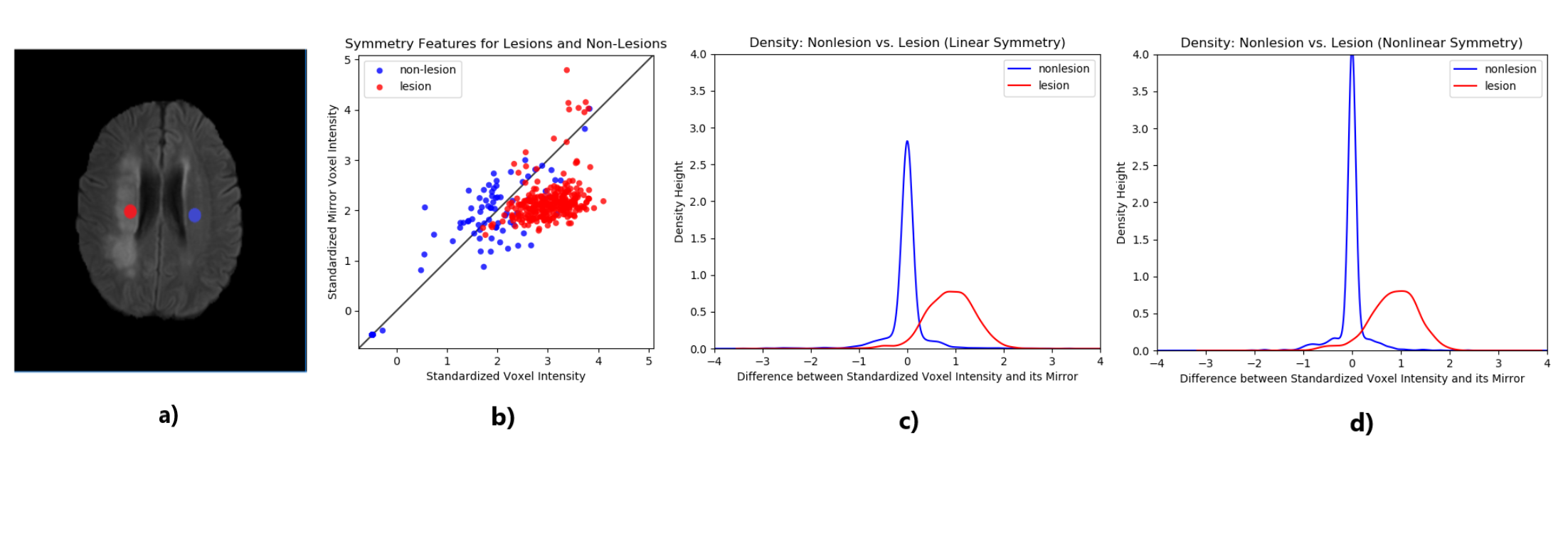}}
	\caption{The quasi-symmetry property of the normal brain can be used to aid lesion segmentation.
Subfigure (a) shows a lesion voxel (red) and its non-lesion (blue) mirror voxel (projected onto the same axial slice); (b) shows the voxel intensity plotted against the intensity of its mirror for a sample of 600 voxels taken from the same brain and equally divided between lesions and non-lesions; (c) and (d) show superimposed densities for the difference between standardized voxel intensity and its mirror based on a sample of 2000 voxels taken from the same brain and equally divided amongst lesions and non-lesions using affine and nonlinear reflective registration respectively. } \label{fig:quasi}
\end{figure}

We propose improvements to existing segmentation methods that exploit the bilateral quasi-symmetry of healthy brains, which breaks down when lesions are present. 
The basic idea is illustrated in Figure \ref{fig:quasi}. The first subfigure shows an axial slice of a brain with a lesion in one hemisphere, and two homologous (``mirror'') voxels, i.e voxels in corresponding parts of the brain but in opposite hemispheres.
In healthy normal brains, there is a strong correlation between intensities of homologous voxels. In lesioned brains, voxels
in a lesion often have intensities very different from the intensities of their homologous voxels, as shown in Fig. \ref{fig:quasi} (b). Note that lesions are typically restricted to one hemisphere, so the homologous voxel of a lesion voxel is almost always non-lesion, which often results in large intensity differences.
On the other hand, lesions typically represent a small proportion of total brain volume, so non-lesion voxels typically have
non-lesion mirror voxels as well, typically resulting in small intensity differences (if the mirror voxels have been accurately located).
The distribution of these intensity differences is illustrated in Fig. \ref{fig:quasi} (c) and (d), for lesion and non-lesion voxels, using affine and nonlinear registration respectively. The increase in mass around zero for non-lesions from using nonlinear registration in comparison to affine registration suggests the superiority of the method in locating mirror voxels.
This pattern of intensity differences can be used to aid the classification of a voxel as lesion or non-lesion.
This method is inspired by the clinical practice of radiologists, who make frequent use of comparisons with homologous areas
to detect abnormalities.

Our method, explained in more detail below, uses 3D nonlinear registration of a neuroimage with a reflected version of itself to determine for each voxel its homologous voxel in the other hemisphere. A patch around the homologous voxel is added as a set of new features to the segmentation algorithm. 
To evaluate this method, we implemented two baseline multimodal MRI stroke lesion segmentation algorithms, both based on 2D CNNs,
following Havaei et al.  \cite{Havaei2017brain} and Kamnitsas et al. \cite{kamnitsas2017efficient}, and then augmented them by adding extra symmetry features as described above.
For each architecture, we evaluated the baseline method and two versions of symmetry augmentation: one using affine registration only, and one 
using nonlinear registration.
We compared the performance of these three segmentation methods on the SISS Training dataset of the Ischemic Stroke Lesion Segmentation Challenge (ISLES 2015)\cite{maier2017isles}.  
Though our experiments use 2D CNNs,
our method can be applied without modification to 3D CNNs.

We are aware of two prior works that have also used brain quasi-symmetry to improve the performance of CNN based methods: 
Shen et al. \cite{shen2017efficient}, and Wang et. al. \cite{wang2016deep}.  
Shen et al. use the SIFT-based method of Loy and Eklundh \cite{loy2006detecting} to identify homologous voxels, 
and report a mean
improvement of 3\% in Dice scores on the high-grade (HG) BRATS2013 Training set.
Wang et al. report a higher increase in mean Dice scores, from $0.63$ to $0.78$, on a private test set of 8 brains with chronic stroke lesions. However the method they use is unclear; the absence of such an explanation suggests a simple affine transformation,
perhaps a reflection in the medial (mid-sagittal) plane.
Both groups require homologous voxels to be in the same axial plane, a restriction that our method does not have. 

We note that the idea of using symmetry also appeared in early literature, prior to the widespread use of neural networks, for example in Meier et al.\cite{meier2014appearance}, Schmidt et al.\cite{schmidt2005segmenting}, and Dvorak et al.\cite{dvorak2013automated}. These works are based on an initial affine registration of each subject's brain to a template. Tustison et al.\cite{tustison2015optimal} also relies on a template, with the major differences being the use of multiple modalities and nonlinear registration. 
Our contributions are thus two-fold:(1) Using template-free registration of an image with a reflected version of itself, called reflective registration, and (2) demonstrating nonlinear reflective registration is better than linear reflective registration for locating mirror voxels, therby improving segmentation.      

\section{Methods}

\begin{figure} 
\centerline{\includegraphics[scale=0.25]{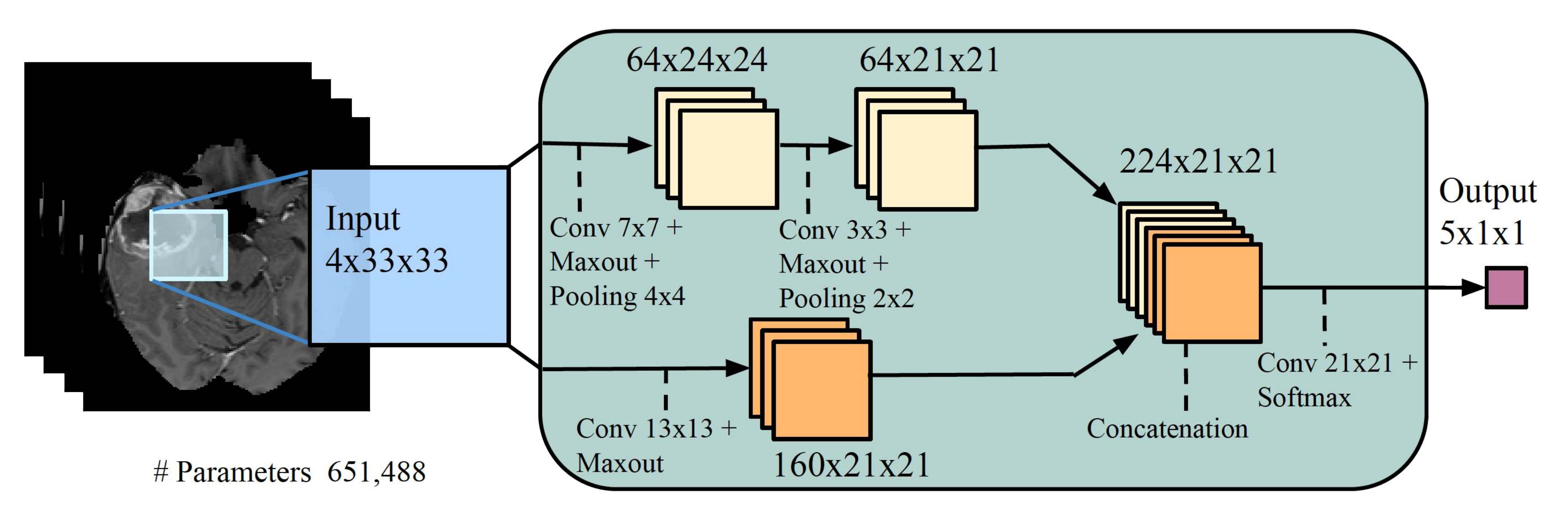}}
\caption{TwoPathCNN architecture, reproduced from Havaei et al. \cite{Havaei2017brain} with permission of the authors. Note that the final output is a $2 \times 1 \times 1$ tensor for our application.}
\centering
\label{fig:arch}
\end{figure}

Our two baseline algorithms are slight modifications of: (i) TwoPathCNN by Havaei et al. \cite{Havaei2017brain},
see Figure \ref{fig:arch}; and (ii) Wider2dSeg by Kamnitsas et al. \cite{kamnitsas2017efficient}. All of the 2D architectures in Kamnitsas et al. \cite{kamnitsas2017efficient}, including Wider2dSeg, are two dimensional variants of their 3D deepMedic architecture. The 2D architectures vary in the number of layers, feature maps (FMs) per layer, and FMs per hidden layer. See Table B.1 in Appendix B of Kamnitsas et al. \cite{kamnitsas2017efficient} for more details. We first describe the architectures and training of these baseline models, and then
describe how to compute and append symmetric features so as to preserve dense inference on 2D images of arbitrary size.   

\subsection{Baseline models}
TwoPathCNN and Wider2dSeg are convolutional neural networks. Both architectures take as input one or, as in the case of Kamnitsas' 2D architectures, two stacks of four patches from different MRI modalities. The networks branch into two pathways. TwoPathCNN consists of three convolutional blocks of sizes
shown in Fig. \ref{fig:arch}, which also shows the locations of maxout and max pooling operations. Wider2dSeg is a deeper architecture with 16 convolutional blocks that makes use of multiscale features through downsampling, convolution and upsampling back to the original scale. This allows for a larger area of information to be used. For more details on the architectures, see \cite{Havaei2017brain} and \cite{kamnitsas2017efficient}. An important feature of these architectures is that all the layers of the network are convolutional, enabling dense inference on full images or image segments.

\paragraph{Training:}
We interpret the output of the CNNs as predicted label probabilities, 
and define a training loss function consisting of the negative log likelihood with both L1 and L2 regularization.
This loss is minimised by following a stochastic gradient descent approach on randomly selected minibatches of patches within each brain.

Performance of CNNs depends greatly on the distribution of the training samples. A commonly used approach is to train a classifier on the same number of image patches from each of the classes per minibatch. However since the classes are imbalanced, this approach biases the classifier towards making false positive predictions. 

In TwoPathCNN, we follow the two-phase training proposed in [3] with minibatches of one labeled sample per training instance. In the first phase, the mini-batches are equally divided among lesions and backround. In the second phase, we keep the weights of all layers fixed and retrain only the final layer on patches uniformly extracted to be closer to the true data distribution. 

For Wider2dSeg, the patch size fed down each pathway is larger than the network's receptive field \cite{kamnitsas2017efficient}. This technique, called dense training, increases the effective batch size by constructing mini-batches of more than one labeled sample per training instance since it allows the network to segment a neighborhood of voxels sorrounding the central voxel.  \cite{kamnitsas2017efficient}. This makes the patch size, also called image segment size, an important parameter to tune since larger patch sizes capture more background voxels than smaller patch sizes.  


\subsection{Implementation details:}
We implemented the models using Tensorflow \cite{abadi2016tensorflow}. 
We apply only minimal pre-processing: we normalize within each input channel by subtracting its mean and dividing by its standard deviation. Unlike \cite{Havaei2017brain}, we did not use N41TK bias correction and we did not remove the $1\%$ highest and lowest intensities. Similarily we didn't use batch normalization as proposed in \cite{kamnitsas2017efficient} since it is more of a requirement for 3D architectures. We use standard momentum and fix the momentum coefficient $\mu= 0.6$ throughout training for all architectures. 

\paragraph{TwoPathCNN:}
Weights are initialized from a uniform distribution  on $(-0.005, 0.005)$, as in \cite{Havaei2017brain}, and biases are initialized to zero. Every epoch consists of 10,000 iterations of stochastic gradient descent with momentum on mini-batches of 10 labeled samples. Each sample consists of 4 stacked patches of size $33\times33$, each patch correspond to a different MRI modality, and a label, which is a ground truth label of the central voxel in the patch.
The first phase of training consists of 50,000 iterations or 5 epochs. The minibatches at this stage contain equal number of positive and negative examples.
The learning rate is set to 0.001 decays by a factor of 0.1 \cite{Havaei2017brain} starting from the third epoch.
The second phase of training consists of another 4 epochs of 10,000 iterations each. The minibatches at this stage have the property that approximately 2$\%$ of samples presented in them are labeled as negative. The learning rate is reset to 0.001 and decays by a factor of 0.1 after each epoch. Thus, in total, the model is trained on 900,000 samples. 
The ${\rm{L_1}}$ regularization constant is $10^{-6}$ and the ${\rm{L_2}}$ regularization constant is $10^{-4}$. 
For further regularization, dropout at a rate of 0.5 was applied on hidden layers of the local pathway. 
In all of these implementation details, we follow \cite{Havaei2017brain}, except that the
regularization constants were inspired from \cite{kamnitsas2017efficient} (which used the same dataset as we do), but changed to account for the increased number of parameters. 

\paragraph{Wider2dSeg:} As in \cite{kamnitsas2017efficient}, we use the weight initiliazation method of He et al. \cite{he2015delving}, since deeper architectures are prone to larger signal variance. The bias terms are initialized to zero. We used the RMSProp optimizer for a total of 80,000 iterations with a learning rate of 0.001 and decayed it by a factor of 0.5 \cite{kamnitsas2017efficient} at the the following iterations: 25,000, 39,000, 49,000, 59,000, 71,000, and 75,000. In contrast to \cite{kamnitsas2017efficient}, we use an image segment size of 43 for the first pathway and 75 for the second pathway, which segments the $27^2$ neighborhood around the central voxel per training instance. Mini-batches are of size 12 and equally divided amongst lesions and background. The batch and image segment sizes were chosen to achieve the same effective batch size shown in Table B.1 of Appendix B from \cite{kamnitsas2017efficient}. To regularize the network we follow \cite{kamnitsas2017efficient} and set the ${\rm{L_1}}$ constant to $10^{-8}$, the ${\rm{L_2}}$ constant to $10^{-6}$, and apply dropout at a rate of 0.5 on the last two hidden layers.

\subsection{Symmetry-augmented methods (LSymm and NLSymm)}

For each subject, we augment the four image modalities by four ``mirror'' images produced as follows.
We begin by producing a reflected Flair image by reversing the orientation of the $x$ (left-right) axis, using the fslorient tool 
of FSL \cite{smith2001fsl}. Since the original images are linearly co-registered, this step is approximately a
reflection in the median, i.e. mid-sagittal, plane.
(Flair was chosen due to its ubiquitous use in lesion segmentation, however we intend in later work compare the use of T1
or multiple modalities  in this step.)
We align the result with the original Flair image 
using either affine/``linear'' (LSymm) or nonlinear (NLSymm) registration.
This step uses the SynQuick method in the ANTs package \cite{Avants2009ants}. 
For LSymm, the ``-t a'' option was used, giving a 2-stage rigid+affine registration.
For NLSymm, the default options were used, giving a 3-stage rigid + affine + nonlinear (``SyN'') registration.
In either case, the resulting transformation, composed with a reflection, produces a symmetry transformation 
$T(x,y,z)$ that associates to each voxel its a corresponding a ``mirror'' voxel in the opposite hemisphere.

Once we have obtained our linear or nonlinear symmetry transformation for each subject,
we use it to construct a
Symmetry Difference Image for each modality, by subtracting from each voxel's standardized intensity the intensity of the standardized ``mirror'' voxel,
$S_{r}(x,y,z) = I_{r}(x,y,z) - I_{r}(T(x,y,z))$.

This results in 4 Symmetry Difference Images (SDIs), one for each modality that we use to augment the original 4 images.
For instance in the baseline TwoPathCNN model, for each voxel, one $33\times33$ patch is extracted from each of the 4 MR images and combined into a
$4 \times 33 \times 33$ tensor; in LSymm and NLSymm, one $33\times33$ patch is extracted from each of the 8 MR images (originals plus SDIs) and combined into a
$8 \times 33 \times 33$ tensor.
This double-size tensor is fed into both the local and global pathway of the TwoPathCNN architecture. 
Apart from doubling the number of images from 4 to 8, all details of architecture and training are exactly as in the baseline methods.

\section{Experimental Results}

\subsection{Dataset}
We evaluated our methods on the ISLES2015  (SISS) training data. The training data consists of FLAIR, DWI ,T1 and T1-contrast images of size $230 \times 230 \times 154$, for each of 28 patients with sub-acute ischemic stroke lesions. 
All images are skull-stripped and have isotropic $1mm^3$ voxel resolution. 

\subsection{Experiment}
For each architecture, we compare the three methods described above: baseline, baseline with LSymm, and baseline with NLSymm; using 
7-fold cross-validation on the Training dataset of 28 subjects. All methods are run with the same hyperparameters, on the same pseudo-random sequence of training patches. We ran our code on a GPU, which took 6 hours to complete one training session, and 30 seconds to segment an entire brain from the validation set.

\begin{figure}[h]
	\centerline{\includegraphics[scale=0.31]{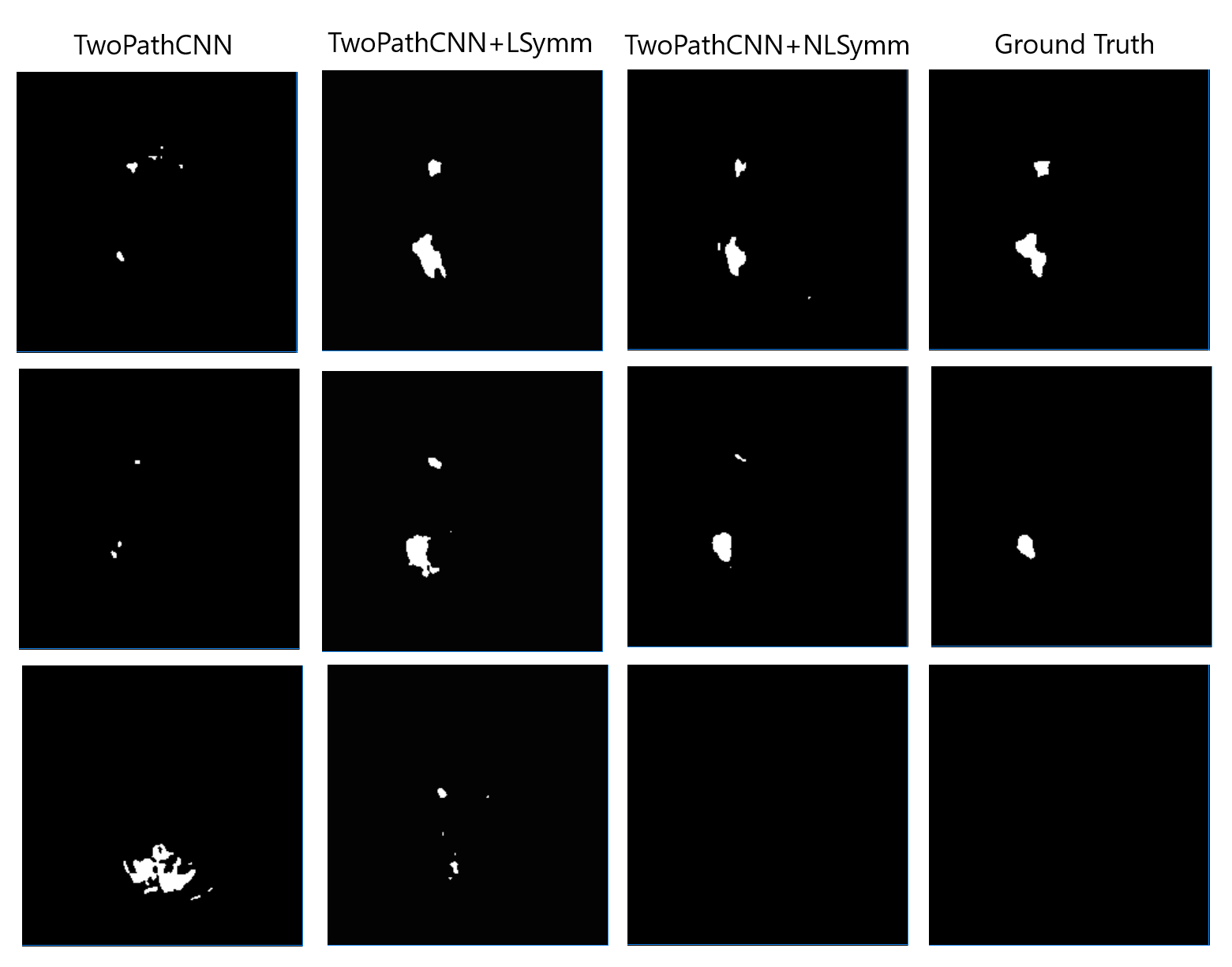}}
	\caption{An example of different segmentations of Brain \#8 of the SISS2015 Training set from the ISLES2015 Challenge. From left to right, the columns show segmentations produced by the TwoPathCNN, TwoPathCNN+LSymm and TwoPathCNN+NLSymm methods, and the ground truth. 
	}
	\label{fig:segex}
\end{figure}

\subsection{Results}


The main results are summarised in Table 1. 
Adding linearly or nonlinearly registered symmetry features to the baseline architectures consistently improves mean Dice coefficient, Recall and Precision showing the efficiency of reflective registration. Wider2dSeg seems to benefit significantly more from the symmetry augmentation, perhaps due to its deeper architecture. Example segmentations produced
by the three methods on TwoPathCNN are shown in Fig. \ref{fig:segex}.

\begin{table}[h!]
	\centering
	\caption{Performance of TwoPathCNN and Wider2dSeg based on a 7-fold cross-validation for baseline, NLSymm and LSymm on the ISLES2015 (SISS) training data. Results for Dice, Recall and Precision are shown as mean (std. dev.). 
	}\label{tab1}
	\begin{tabular}{|c|c|c|c|}
		\hline
		\bfseries Architecture &  \bfseries Dice & \bfseries Recall & \bfseries Precision\\ 
		\hline
		TwoPathCNN &  0.45(0.25) &  0.59(0.22) &  0.45(0.29)\\
		\hline
		TwoPathCNN+LSymm & 0.52(0.23) & 0.63(0.23) & 0.50(0.28)\\
		\hline
		TwoPathCNN+NLSymm & \bfseries 0.54(0.21) &  \bfseries 0.65(0.22) &  \bfseries 0.52(0.26)\\
		\hline
		Wider2dSeg &  0.49(0.25) &  0.53(0.28) &  0.54(0.25)\\
		\hline
		Wider2dSeg+LSymm & 0.61(0.22) & 0.58(0.25) & 0.67(0.22)  \\
		\hline
		Wider2dSeg+NLSymm & \bfseries 0.62(0.22) & \bfseries 0.60(0.25) & \bfseries 0.68(0.21) \\

		\hline
	\end{tabular}
\end{table}
 
Moreover, nonlinearly registered symmetry features consistently produces higher Dice, Recall and Precision scores compared to linearly registered symmetry features.

\section{Discussion}

We have proposed an improvement to existing segmentation methods by exploiting the bilateral quasi-symmetry of healthy brains, without the need for a template, which breaks down when lesions are present.
The method consists of augmenting the input images to a CNN with extra Symmetry Difference Images 
consisting of intensity differences between homologous (``mirror'') voxels in different hemispheres.
We showed how to incorporate these symmetric features into the increasingly popular patch-based CNNs so as to preserve dense inference. In a comparison on the ISLES2015 SISS datset, we found that adding symmetric features 
generated using nonlinear reflective registration (NLSymm method) consistently resulted in a mean improvement in the Dice coefficient.
Using linear reflective registration instead gave consistently smaller improvements over baseline showing that nonlinear registration 
is superior in this application.

We have shown that the brain's quasi-symmetry property is a valuable tool for brain lesion segmentation. 
The ease of application of symmetry augmentation to most existing CNN methods suggest a 
potentially wide-ranging utility of the method.


%
%
%
%

 \bibliographystyle{splncs04}
 \bibliography{Raina2019arXiv}
\end{document}